# Optimal Design of Experiments for Simulation-Based Inference of Mechanistic Acyclic Biological Networks

**Vincent D. Zaballa**
Department of Biomedical Engineering
University of California, Irvine
Irvine, CA 92687
vzaballa@uci.edu

**Elliot E. Hui**
Department of Biomedical Engineering
University of California, Irvine
Irvine, CA 92687
eehui@uci.edu

## Abstract

Biological signaling pathways based upon proteins binding to one another to relay a signal for genetic expression, such as the Bone Morphogenetic Protein (BMP) signaling pathway, can be modeled by mass action kinetics and conservation laws that result in non-closed form polynomial equations. Accurately determining parameters of biological pathways that represent physically relevant features, such as binding affinity of proteins and their associated uncertainty, presents a challenge for biological models lacking an explicit likelihood function. Additionally, parameterizing non-closed form biological models requires copious amounts of data from expensive perturbation-response experiments to fit model parameters. We present an algorithm (SBIDOEMAN) for determining optimal experiments and parameters of systems biology models with implicit likelihoods. We evaluate our algorithm using simulations of held-out true parameter values and demonstrate an improvement in the rate of accurate parameter inference over random and equidistant experimental designs when evaluated on two simple models of the BMP signaling pathway with an implicit likelihood function.

## 1 Introduction & Related Work

Systems biology, the modeling and study of complex biological systems by dynamical models, seeks to understand mechanisms of individual parts by studying the whole system. These systems are usually modeled by Ordinary Differential Equations (ODEs) that model the biology of proteins binding to one another or reactions occurring within a cell. Studying the response of the system can be used to gain an understanding of latent processes underway when a cell responds to perturbations in its environments. Understanding dynamical systems of cells and how they respond to perturbations is important in drug design, where misunderstanding can lead to reduced drug efficacy and increased off-target effects.

However, dynamical systems constrained by physics and experimental limitations, such as measuring a single time point after perturbing a system using flow cytometry, can lead to polynomials with non-closed form steady-state solutions that do not admit an explicit likelihood function. For example, the steady state solution for a model of Bone Morphogenetic Protein (BMP) ligands binding to BMP receptors and then sending a downstream gene expression signal can be solved by least squares regression [1] or convex optimization [2]. While these methods provide a solution, they do not admit an explicit likelihood function that can be used directly with methods such as Markov Chain Monte Carlo (MCMC) to determine model parameters and their uncertainty. In this case, the model of BMP binding has an *implicit* likelihood function, which is an unknown or intractable likelihood of the data, and also known as a generative model. This is a common scenario in biology, where certain systems can be simulated but do not have an explicit likelihood function, such as systems of stochastic biological functions [3] and metabolic pathways [4].

Traditional approaches to determining the parameters of a model with an implicit likelihood used Approximate Bayesian Computation (ABC) techniques, akin to guessing parameters a simulator may need to return the observed data and accepting those parameters that fall within a user-specified distance. However, this technique is slow and also typically dependent on user-defined summary statistics of the observed data, $x_o$ [3, 5].

Recent likelihood free inference (LFI) methods based on neural networks that estimate the density, or probability distribution, of each unknown parameter, $\theta$, have shown to improve performance over classic ABC methods [6]. LFI methods, also known as simulation-based inference (SBI), were recently benchmarked on various tasks and settings, and demonstrated reliably more efficient and effective in estimating parameters than ABC methods across a range of tasks [7].

Determining the parameters that may describe the biological system given experimental designs, $p(\theta|x_o)$, is important, but it is also important to design experiments to arrive at an accurate parameterization with the least number of experiments. Recent work has applied optimal experimental design to perturbation experiments to study hematopoetic stem cell (HSCs) systems [8] and chemical design and synthesis [9], but there lack methods applied to perturbation-response biological settings, where the goal is understanding dynamical biological systems, such as dosing cells in microwell plates and measuring their response after an incubation period. Using uncertainty estimates, or entropy, and information-based objective functions [10], optimal experiments can be designed to determine parameters of dynamical systems by LFI given a model of the dynamical system, its parameter priors, and observed data.

We propose and test an algorithm for implicit biological systems that:

- Determines the parameters and their uncertainty using LFI.
- Uses uncertainty information to design new experiments.
- Performs better than controls when benchmarked on two implicit models of the BMP signaling pathway.

Accurate parameterizations of biological systems is an ongoing area of research that has resulted in methods such as graph-based models enclosed in an activation function to parameterize models of systems biology [11]. While previous methods may be effective at parameterizing a set of known biological connections and predicting responses to perturbation, these methods lack an uncertainty estimate that can be used to determine experiments that maximize the mutual information between prior model parameters and predictive posteriors given proposed experimental designs. Previous work has applied ABC methods to systems biology [3]; we see our work as an extension of LFI methods in systems biology that simultaneously harnesses entropy for optimal experimental designs.

## 2  Background

**Normalizing Flows.** Normalizing flows are a class of invertible and differentiable neural networks that describe a series of monotonic functions that can either minimize the divergence of the pushforward from a base distribution, $p_u(\mathbf{u})$, which is typically a Gaussian distribution, to the data $p_X(\mathbf{x})$, or vice versa via a pullback. Formally, we use the change of variable formula and a composition of monotoic diffeomorphic functions, $\mathbf{f}_\phi$, which can be neural networks parameterized by $\phi$, to transform data from a base distribution, $p_u(\mathbf{u})$, to the data distribution, $p_X(\mathbf{x})$, as follows:

$$p_X(\mathbf{x}) = p_u(f_\phi^{-1}(\mathbf{x})) \left| \det \frac{\partial \mathbf{f}_\phi^{-1}}{\partial \mathbf{x}} \right|. \tag{1}$$

See Papamakarios et al. for details on the theory and implementation of normalizing flows.

**SBI.** In parallel to recent innovations normalizing flow architectures, much work has focused on algorithms for sequential posterior estimation by estimating the posterior [6, 13], likelihood [14], and ratios of posteriors to priors [15] to estimate the posterior $p(\theta|x_o)$ of a model of interest given observed data $x_o$. SBI methods are used extensively in fields where functions can be simulated but not evaluated, such as particle physics [16]. The SBI method used in this paper is known as Sequential Neural Posterior Estimation (SNPE), which uses a neural network to directly estimate the posterior distribution [13]. SNPE aims to estimate the posterior directly, $\bar{q}_{x,\phi}$, by

$$\bar{q}_{x,\phi} = q_{F(x,\phi)}(\theta) \frac{\tilde{p}(\theta)}{p(\theta)} \frac{1}{Z(x,\phi)}, \qquad (2)$$

where $q_{F(x,\phi)}(\theta)$ is a normalizing flow that estimates the posterior $p(\theta|x)$, $Z(x,\phi)$ is a normalization constant, and $\tilde{p}(\theta)/p(\theta)$ is a user-defined importance weighting factor. See Greenberg, Nonnenmacher, and Macke for more details.

**Design of Experiments (DOE) for Implicit Models.** While much recent research has focused on developing novel normalizing flow and SBI methods, DOE for models with implicit likelihoods has only recently seen increased attention, with a focus on evaluating different score functions of estimates of the mutual information's lower and upper bounds between a model's priors and predictive posterior [17, 18]. Commonly, most methods start by finding the optimal experimental design, $\mathbf{d}^*$ that maximizes a utility function, $U(\mathbf{d})$, describing the change in entropy of model parameters before and after an experiment with design $\mathbf{d}$ is conducted. This optimization problem is described as

$$\mathbf{d}^* = \underset{d \in \mathcal{D}}{\mathrm{argmax}}\, U(\mathbf{d}), \qquad (3)$$

where $\mathcal{D}$ represents the space of feasible designs. The utility function can then be formulated as the mutual information, $I(\theta, y|d)$ between $\theta$ and $y$ given a certain design $d$,

$$U(\mathbf{d}) = \mathbf{I}(\theta, y|d) = \mathbb{E}_{p(\theta)p(y|\theta,d)}\left[\log \frac{p(y|\theta,d)}{p(y|d)}\right], \qquad (4)$$

which results in the expected information gain given a certain experiment, $d$. Various upper and lower bound of the mutual information have been proposed [17, 19]. We use an estimate of the lower bound of the mutual information using the Donsker-Varadhan lower bound calculated by a Mutual Information Neural Estimation (MINE) network [20]. This lower bound is then used as the objective function of a Gaussian process within a Bayesian Optimization routine [18].

Altogether, these parts constitute the Simulation-Based Inference Design Of Experiment for biological Mechanistic Acyclic Networks (SBIDOEMAN) algorithm (Appendix A).

## 3 Results

We evaluated how SBIDOEMAN performed on two simple models of the BMP pathway, called the onestep and twostep models, (Appendix B) with held-out parameters representing the binding affinity and phosphorylation efficiency of physically-relevant variables in the BMP model. We compared the SBIDOEMAN algorithm to random experimental designs and log-equidistant titrations of ligands from $10^{-3}$ to $10^3$ ng/mL of BMP ligand as a design with a budget of 5 experimental designs for each condition. The same SNPE-based SBI with neural spline flow (NSF) [21] normalizing flow was used for each experimental design policy tested. For each model, we trained an ensemble of independent SNPE density estimators with a sample size varying form 38 to 50 completed inferences given a time budget of 8 hours to complete. Using independent ensembles helped determine a distribution of reported metrics and was a valuable tool for debugging SBIDOEMAN (Appendix C).

We compared the performance by the root mean squared error (RMSE) discrepancy between the maximum a posteriori (MAP) point estimate of the inferred posterior distribution, $p(\theta|x_o)$ and known true parameter values, $\theta_T$. The results of SBIDOEMAN on the onestep and twostep models are shown in Table 1. The SBIDOEMAN outperformed each control policy using a RMSE metric. To gain a better understanding of the difference in policy between SBIDOEMAN and random search, we examined violin plots representing the posterior distribution of an ensemble of distributions representing the RMSE of the MAP estimate over the 5 designs, as shown in Fig. 1. The improvement in policy compared to the random search is clear in the simpler onestep BMP model, where random search has wider variance after the initial design, and subtly shows in the more complicated twostep BMP model in the last design.

|                | Policy                          |                               |                               |
|----------------|---------------------------------|-------------------------------|-------------------------------|
| BMP Model Type | SBIDOEMAN                       | Random                        | Equidistant                   |
| Onestep        | **0.004 ± 0.007** ($n=48$)      | $0.013 \pm 0.035$ ($n=38$)    | $0.023 \pm 0.051$ ($n=50$)    |
| Twostep        | **0.149 ± 0.153** ($n=48$)      | $0.242 \pm 0.146$ ($n=40$)    | $0.249 \pm 0.173$ ($n=50$)    |

Table 1: Mean and standard error of RMSE of an ensemble of MAP estimate of the posterior compared to true held-out parameter values after 5 sequential experimental evaluations of SBIDOEMAN compared to random search and equidistant controls. Lower RMSE is better. The number of samples vary due to rejection sampling from the posterior surpassing the 8-hour allocated simulation budget. Results indicate that for two models of the BMP pathway, SBIDOEMAN was able to perform an order of magnitude better than random and equidistant search with no, or minimal, overlap of standard errors for the onestep model, and better for the twostep model.

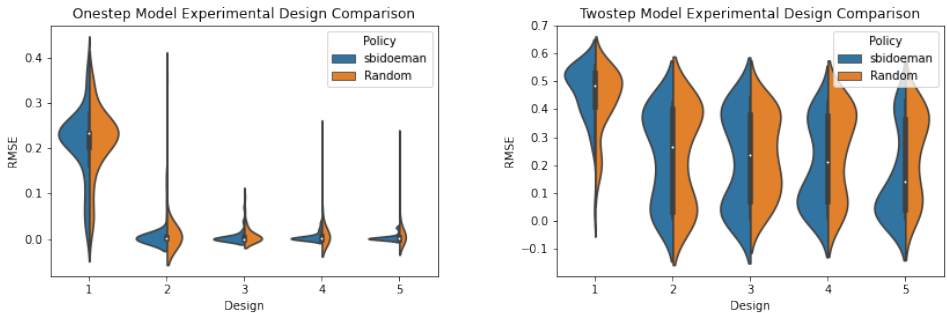

Figure 1: Comparing search policy of the SBIDOEMAN and random search across an ensemble of models shows an improvement in the convergence of the SBIDOEMAN to the true value with less variance for both onestep (left) and twostep (right) models. For the onestep model, a simpler model with only two unknown parameters, SBIDOEMAN arrives at an accurate MAP estimate of the true parameter values with RMSE of $0.01 \pm 0.03$ in just 2 designs. When examining the difference between experimental design policies in the twostep model, which has 3 unknown parameters, SBIDOEMAN showed more gradual improvement over random search to arrive at a lower RMSE MAP estimate of the correct held-out parameter values. However, improvement can qualitatively be seen by the last design, indicating that more designs may be required for more complex models to converge but that SBIDOEMAN is more efficient at arriving at true parameter values than random search.

## 4  Discussion

We demonstrated using the SBIDOEMAN algorithm to both estimate parameters of a biophysical model with an intractable likelihood and design optimal experiments to gain more information than using a sub-optimal search strategy. We compared to random search as a baseline, and equidistant dosing, which is common when evaluating Hill Functions of titration curves during drug screening. Our method shows an improvement in the rate and accuracy of parameterizing implicit biological functions over an equidistant method. This improvement is critical whenever samples are scarce, such as assessing drug combinations on cancer biopsies.

Future work will evaluate the effectiveness of the SBIDOEMAN algorithm on experimental data. Although neural network-based SBI methods demonstrate improvements over ABC methods, they are still computationally expensive and complex in the number of hyperparameters to assess. However, recent works have shown the ability to include symmetries of data in normalizing flow architectures [22, 23, 24]. Given that breaking Lie symmetries is a strategy for determining differential equations [25, 26], this may help in designing experiments where multiple models are candidates to represent the true underlying biology, such as whether homodimeric and heterodimeric BMP ligands operate by different models [27], and potentially reduce the computational burden and increase the utility of normalizing flows for experimental design and model selection in systems biology.

## Acknowledgements

This research was funded by the National Institute of General Medical Sciences (NIGMS) of the National Institutes of Health (NIH) under award number R01GM134418. We would like to thank the Elowitz Lab for helpful discussions and Christina Su for making the simulator that made this work possible.## Acknowledgements

This research was funded by the National Institute of General Medical Sciences (NIGMS) of the National Institutes of Health (NIH) under award number R01GM134418. We would like to thank the Elowitz Lab for helpful discussions and Christina Su for making the simulator that made this work possible.

## A  The SBIDOEMAN Algorithm & Choice of Hyperparameters

The SBIDOEMAN algorithm is described using a simulator of the BMP model as a surrogate for collecting experimental data. When experimentally validating the algorithm, the experimental collection process needs to be replaced by an iterative experimental process. Potentially confusing nomenclature for the SBIDOEMAN algorithm is the difference between the number of SBI rounds, $N_S$, which is the number of rounds of posterior refinement in the SBI module, and the number of experimental rounds, $N_E$, which is the total number of experiments to perform.

---

**Algorithm 1** SBIDOEMAN

1: **Require:** Simulator of an implicit model, $f(d, \theta)$, which accepts experimental designs $d$ and parameters $\theta$, held-out true parameters $\theta_T$, number of simulations per SBI round $N_S$, number of rounds of SBI $N_R$, number of experiments $N_E$, choice of neural density estimator $q_\phi(x|\theta)$, number of prior samples to use for MINEBED $n$, and priors over parameters $p(\theta)$
2: **Return:** Approximate posterior $p(\theta|x_o, d)$, estimated optimal designs $d$, observed data $x_o$
3: Initialize a design $\mathbf{d_0}$ by random sampling and set $d = d_0$
4: Initialize MINE neural network parameters $\boldsymbol{\psi}_0$
5: Set $\tilde{p}_1(\theta) \coloneqq p(\theta)$
6: **for** $i = 1 : N_E$ **do**
7:     Draw $n$ samples from the prior distribution of model parameters $\boldsymbol{\theta} : \boldsymbol{\theta}^{(1)}, \dots, \boldsymbol{\theta}^{(n)} \sim \tilde{p}_i(\boldsymbol{\theta})$
8:     Simulate data $x^{(i)}, i = 1, \dots, n$ using current design, $d$, and prior simulations, $\boldsymbol{\theta}$, using the provided simulator $f(d, \boldsymbol{\theta})$
9:     Select $d^*$ using MINEBED as shown in Equation 4 by gradient ascent of MINE neural network parameters, $\psi$, and Bayesian optimization of resulting lower bound measure of $\widehat{I}(\theta, y; d)$ using a Gaussian process to select $d^*$
10:     Perform experiment using $d^*$ and observe experimental condition $x_o = f(d^*, \theta_T)$
11:     **for** $j = 1 : N_R$ **do**
12:         **for** $k = 1 : N_S$ **do**
13:             Sample $\theta_{j,k} \sim \tilde{p}_j(\theta)$
14:             Simulate $x_{j,k} \sim f(x, \theta_{j,k})$
15:         **end for**
16:         $\phi \leftarrow \underset{\phi}{\operatorname{argmin}} \sum_{m=1}^{j} \sum_{k=1}^{N_S} -\log \tilde{q}_{x_m,k,\phi}(\theta_{m,k})$ by Equation 2
17:         $\tilde{p}_{j+1}(\theta) \coloneqq q_{F(x_o, \phi)}(\theta)$
18:     **end for**
19:     Set $\tilde{p}_i(\theta) \coloneqq q_{F(x_o, \phi)}(\theta)$
20: **end for**

---

When implementing SBIDOEMAN in code, there are multiple opportunities to reuse samples from different sections of the code in order to amortize sampling, but are omitted here for brevity. The hyperparameters we chose were $N_S = 500$, $N_R = 2$, and a NSF architecture with 150 hidden features (neurons), 10 transforms, and 20 bins.

It will be interesting for future work to include constrained optimization in this algorithm given a finite resource for the number of designs, $d$, for DOE of implicit models. Then, $N_E$ will be the result of the constrained optimization problem.

## B  Physical Models of the BMP Pathway

The BMP signaling pathway can be described by mass action kinetics of proteins binding to one another and conservation laws to describe the process of a downstream genetic expression signal reaching a steady-state based on receptors available and ligands in a cell's environment. Varying degrees of model complexity can be formulated and used to describe observed biological data. The **twostep** model of BMP signaling was originally proposed by Antebi et al. This system is described as $n_L$ ligands, $L_j$, binding to one of $n_A$ Type $A_i$ receptor to form a heterodimeric complex, $D_{ij}$, which then binds to one of $n_B$ type $B_k$ receptors to form a trimeric complex, $T_{ijk}$. An assumption made was that the reactions are reversible with forward rates $k^D_{f_{ij}}$ and $k^T_{f_{ij}}$ for dimeric and trimeric complex formation, and $k^D_{r_{ij}}$ and $k^T_{r_{ij}}$ for the reverse reaction rates. This model's chemical equilibrium equations are

$$A_i + L_j \underset{K^D_{r_{ij}}}{\overset{K^D_{f_{ij}}}{\rightleftharpoons}} D_{ij} \tag{5}$$

$$D_{ij} + B_k \underset{K^T_{r_{ijk}}}{\overset{K^T_{f_{ij}}}{\rightleftharpoons}} T_{ijk}, \tag{6}$$

where there is a chemical equilibrium between the formation of a Dimeric Ligand-receptor complex and trimeric complex and its respective dimeric and type B receptor.

The twostep was followed by a simpler model by Su et al. called the **onestep** model, modeling only one step to form the Trimeric complex of Ligand, Type A, and Type B receptors:

$$A_i + B_k + L_j \underset{K_{r_{ijk}}}{\overset{K_{f_{ijk}}}{\rightleftharpoons}} T_{ijk}. \tag{7}$$

Relevant to this paper, the onestep model uses one less binding affinity to model the rate of downstream signal expression than the twostep model.

Both models found each complex $T_{ijk}$ phosphorylates an intracellular second messenger at a rate $\epsilon_{ijk}$ to generate gene expression signal $S$, which degrades at a rate $\gamma$. This differential equation is shown as:

$$\frac{dS}{dt} = \sum_{j=1}^{n_L} \sum_{i=1}^{n_A} \sum_{k=1}^{n_B} \epsilon_{ijk} T_{ijk} - \gamma S \tag{8}$$

Both onestep and twostep models can be represented by ordinary differential equations (ODEs); however, ODEs do not reflect the experimental constraints in place when modeling the reaction of cells to ligand in a contained volumetric environment where ligands do not degrade. Considering ligands do not degrade and *in vitro* evaluation of cells' response to ligands is measured in a microwell plate with fixed volume, conservation laws turn the ODE into an algebraic system of equations. Under this regime, where volume of ligands is large and there are significantly more ligands than receptors, ligand concentration can be assumed to remain constant. Additionally, by assuming that production and consumption of receptors are in steady state, conservation of mass of each molecule enforces a set of algebraic equations. Letting $L_j^0$, $A_i^0$, and $B_k^0$ represent initial values of each species, for the **onestep** model, we obtain the following constraints:

$$L_j^0 = L_j \tag{9}$$

$$A_i^0 = A_i + \sum_{j=1}^{n_L} \sum_{k=1}^{n_B} T_{ijk} \tag{10}$$

$$B_k^0 = B_k + \sum_{j=1}^{n_L} \sum_{i=1}^{n_A} T_{ijk} \tag{11}$$

The assumption of steady-state equilibrium is made because the binding and unbinding of ligands and receptors occurs at a faster time scale than downstream gene expression. Hence, the time derivatives of any ODEs vanish and the binding affinity, $K_{ijk} \equiv K_{f_{ijk}}/K_{r_{ijk}}$, and phosphorylation efficiency, $\epsilon_{ijk} \equiv \epsilon_{f_{ijk}}/\gamma$ turns into the algebraic equations:

$$T_{ijk} = K_{ijk} L_j A_i B_k \tag{12}$$

$$S = \sum_{i=1}^{n_L} \sum_{j=1}^{n_A} \sum_{k=1}^{n_B} e_{ijk} T_{ijk}. \tag{13}$$

We can rearrange eqs. 10 and 11 by solving for steady-state values of $A_i$ and $B_k$, and combine with eq. 12 to arrive at a system of $n_T = n_L n_A n_B$ quadratic equations for $T_{ijk}$:

$$T_{ijk} = K_{ijk} L_j \left( A_i^0 - \sum_{j'=1}^{n_L} \sum_{k'=1}^{n_B} T_{ij'k'} \right) \left( B_k^0 - \sum_{j'=1}^{n_L} \sum_{i'=1}^{n_A} T_{i'j'k} \right). \tag{14}$$

The solutions for $T_{ijk}$ can be substituted into eq. 13 and solved by least squares regression or convex optimization. However, an explicit solution is not readily available, as solving the equation results in multiple positive, real-valued, discriminant solutions that can be distinguished in simple models by qualitative interpretation of the solutions. Thus, difficulty in determining the discriminant makes this model of BMP signaling an implicit model. Future work looking into solving this problem would alleviate the need to use an implicit model.

## C  Choice of Normalizing Flow

An important choice when conducting SBI is the type of normalizing flow used, where there are tradeoffs between computational complexity and accuracy. A simple neural network that we tested was the Mixture Density Network [29] trained by Stochastic Variational Inference (SVI). This network is easy to sample but not as sensitive to non-Gaussian distributions. Another option we considered are neural spline flows, which are flexible likelihood estimators that are relatively fast to perform inference and sampling. Using an ensemble of neural density estimators can help to evaluate the performance of the choice of normalizing flow for the task at hand. We noticed an improvement in the simple onestep BMP model when switching from a MDN to a NSF, as denoted by the decrease in variance of MAP RMSE over subsequent experimental design rounds and shown in Figure 2.

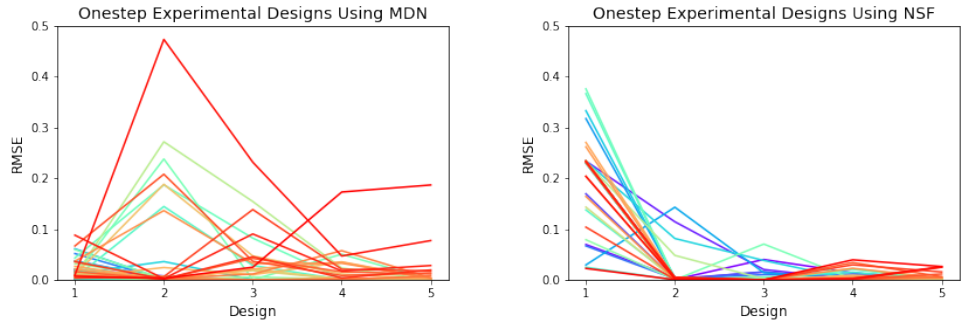

Figure 2: Comparing different normalizing flows by ensembles of SBIDOEMAN. We can see that the MDN architectures (left) had increased variance in RMSE values over experimental runs while the NSF architecture (right) converged more rapidly and with less variance. The color of the lines indicate the ranking of the final RMSE, where red represents the highest RMSE and blue represents the lowest final RMSE.